\documentclass[conference]{IEEEtran}
\IEEEoverridecommandlockouts
\usepackage{cite}
\usepackage{amsmath,amssymb,amsfonts}
\usepackage{algorithmic}
\usepackage{graphicx}
\usepackage{textcomp}
\usepackage{xcolor}
\usepackage{url}
\usepackage{tabularx}
\usepackage{subfig}
\usepackage{comment}

\def\tablename{Table}
\usepackage{etoolbox}
\makeatletter
\patchcmd{\@makecaption}
  {\scshape}
  {}
  {}
  {}
\makeatletter
\patchcmd{\@makecaption}
  {\\}
  {.\ }
  {}
  {}
\makeatother
\def\tablename{Table}
\def\BibTeX{{\rm B\kern-.05em{\sc i\kern-.025em b}\kern-.08em
    T\kern-.1667em\lower.7ex\hbox{E}\kern-.125emX}}
\begin{document}

% The mapping of quantum algorithms in multi-core quantum computing architectures

\title{Mapping quantum algorithms to multi-core quantum computing architectures\\
}

\author{\IEEEauthorblockN{Anabel Ovide\IEEEauthorrefmark{1},
 Santiago Rodrigo\IEEEauthorrefmark{2}, Medina Bandic\IEEEauthorrefmark{3}, Hans Van Someren\IEEEauthorrefmark{3},  Sebastian Feld\IEEEauthorrefmark{3}, Sergi Abadal\IEEEauthorrefmark{2}, \\
 Eduard Alarcon \IEEEauthorrefmark{2}, and Carmen G. Almudever\IEEEauthorrefmark{1}}

\IEEEauthorblockA{\IEEEauthorrefmark{1}\textit{Technical University of Valencia, Spain}}
\IEEEauthorblockA{\IEEEauthorrefmark{2}\textit{Technical University of Catalonia, BarcelonaTech, Spain}}
\IEEEauthorblockA{\IEEEauthorrefmark{3}\textit{Technical University of Delft, The Netherlands}}}

\begin{comment}
\author{\IEEEauthorblockN{1\textsuperscript{st} Given Name Surname}
\IEEEauthorblockA{\textit{dept. name of organization (of Aff.)} \\
\textit{name of organization (of Aff.)}\\
City, Country \\
email address or ORCID}
\and
\IEEEauthorblockN{2\textsuperscript{nd} Given Name Surname}
\IEEEauthorblockA{\textit{dept. name of organization (of Aff.)} \\
\textit{name of organization (of Aff.)}\\
City, Country \\
email address or ORCID}
\and
\IEEEauthorblockN{3\textsuperscript{rd} Given Name Surname}
\IEEEauthorblockA{\textit{dept. name of organization (of Aff.)} \\
\textit{name of organization (of Aff.)}\\
City, Country \\
email address or ORCID}
\and
\IEEEauthorblockN{4\textsuperscript{th} Given Name Surname}
\IEEEauthorblockA{\textit{dept. name of organization (of Aff.)} \\
\textit{name of organization (of Aff.)}\\
City, Country \\
email address or ORCID}
\and
\IEEEauthorblockN{5\textsuperscript{th} Given Name Surname}
\IEEEauthorblockA{\textit{dept. name of organization (of Aff.)} \\
\textit{name of organization (of Aff.)}\\
City, Country \\
email address or ORCID}
\and
\IEEEauthorblockN{6\textsuperscript{th} Given Name Surname}
\IEEEauthorblockA{\textit{dept. name of organization (of Aff.)} \\
\textit{name of organization (of Aff.)}\\
City, Country \\
email address or ORCID}
}
\end{comment}

\maketitle

\begin{abstract}
Current monolithic quantum computer architectures have limited scalability. One promising approach for scaling them up is to use a modular or multi-core architecture, in which different quantum processors (cores) are connected via quantum and classical links. This new architectural design poses new challenges such as the expensive inter-core communication. To reduce these movements when executing a quantum algorithm, an efficient mapping technique is required. In this paper, a detailed critical discussion of the quantum circuit mapping problem for multi-core quantum computing architectures is provided. In addition, we further explore the performance of a mapping method, which is formulated as a partitioning over time graph problem, by performing an architectural scalability analysis.

%So far, there is only one technique for mapping quantum algorithms in multi-core quantum architectures, based on a partitioning over time graph problem. In this paper, we explore its performance by performing an architectural scalability analysis.
\end{abstract}


 \begin{figure*}[!h]
\centering
 \includegraphics[width=\linewidth]{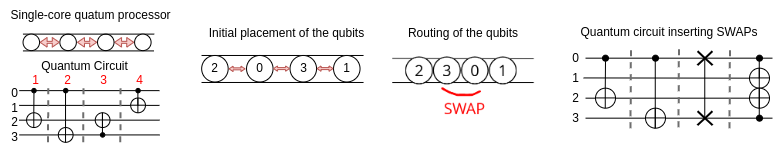}
\caption{
An illustrative example of the quantum circuit mapping procedure. We consider  a 1-D linear array quantum processor shown at the top-left, where only adjacent qubits (circles) can interact. Next, an optimal initial placement of qubits is performed based on the quantum circuit to be executed. Note that, the first two CNOT gates (CNOTS in time steps 1 and 2) can be directly performed as qubits 0 and 2 as well qubits 0 and 3 are adjacent. However, an extra SWAP gate has to be inserted for allowing the execution of the other two CNOT gates between qubit 2 and  qubit 3 and between quit 0 and qubit 1. A SWAP gate exchanges the state of the two involved qubits. In addition, note that scheduling the quantum gates saves one time step as the last two CNOT gates can be performed in parallel.}
\label{fig:map_example}
\end{figure*}

\section{Introduction}
%           ---> !!! FIGURES IN THE MAIN FILE FOR EASY ORDERING !!! <----

\begin{comment}
\textcolor{red}{TODO: 1) Review the concept of distributed quantum computing. 2) Acknowledgement section: QUADRATURE? QIA?. 3) read again the paper and check that everything is clear and correct.}
\end{comment}

Quantum computers are a revolutionary technology that can outperform classical computing in areas such as scientific simulation~\cite{https://doi.org/10.48550/arxiv.1903.10541}, cryptography~\cite{Mavroeidis_2018}, machine learning~\cite{9274431}, search or optimization~\cite{Montanaro_2016}, thanks to the use of quantum mechanics phenomena like superposition and entanglement. Current quantum computing technologies, commonly called NISQ (Noisy Intermediate-Scale Quantum)~\cite{Preskill_2018} devices, are limited in the number of qubits and prone to errors. The most advanced quantum processors consist of a few tens of noisy qubits (e.g. IBM's 433-qubit Osprey processor \cite{Osprey}), meaning that their state can be easily modified due to the interaction with the external environment (decoherence) and that quantum gates and measurements are implemented with imperfect operations. Algorithms for NISQ devices have been developed to leverage their scarce and noisy resources such as Quantum Approximate Optimization Algorithm (QAOA) or Variational Quantum Eigensolver (VQE)~\cite{Bharti_2022}. However, to build a universal fault-tolerant quantum computer and achieve the full computational power these machines will provide, it is necessary to scale them up in a way that the number of qubits is increased without incurring much higher error rates. Therefore, the scalability of quantum computing systems is one of the main challenges the quantum community is currently facing.

Nowadays NISQ computers are implemented as single-chip processors, also referred as single-core quantum processors, in which all qubits are integrated within a single chip. This monolithic  architecture is hardly scalable due to challenges in the control electronics and wiring \cite{sebastiano2020cryo}, an increase of undesired interactions between qubits (i.e. crosstalk)~\cite{Ding_2020} and a decrease of the device uniformity and yield. To overcome these challenges and solve the scaling problem, modular quantum computing architectures have been already proposed for different qubit implementation technologies ~\cite{Monroe_2014, https://doi.org/10.48550/arxiv.2201.08825, https://doi.org/10.48550/arxiv.2201.08861,https://doi.org/10.48550/arxiv.2210.10921}. The main idea is to combine multiple quantum processors and connect them via single control systems, classical communication links and ultimately quantum communication technologies~\cite{QuantumIntranet, IBMRoadmap,IBMRoadmap2025}. We refer to the latter, in which both classical and quantum communication channels are incorporated as multi-core quantum computing architectures. They will allow performing distributed multi-core quantum computing in which a large algorithm consisting of more qubits than there are in a single processor, is partitioned into smaller instances and executed on several quantum chips.

With this novel architectural approach, new challenges emerge as pointed out in \cite{rodrigo2021double} that include: i) the implementation of input/output communication ports for each core (processor) as well as the definition of the ratio of qubits devoted to computation and communication; ii) the development of the technology required for communicating quantum information between chips and corresponding communication protocols; and iii) compilation techniques, including placement and routing of qubits and scheduling of quantum operations, that allow for an efficient distributed multi-core quantum computation, which will be the central topic of this paper.

Executing an algorithm on a NISQ processor, requires to perform some modifications on the corresponding quantum circuit such that all quantum hardware constraints are satisfied. This process of adapting the quantum circuit to the quantum processor restrictions is usually called mapping or transpiling. Whereas several quantum circuit mapping techniques have been proposed for single-core quantum architectures~\cite{https://doi.org/10.48550/arxiv.2007.01000,mappings, 8382253, 7059001, Venturelli_2018,10.1145/3297858.3304075} only recently, the first compilation techniques for mapping quantum algorithms onto connectivity-simplified multi-core quantum 
architectures have been proposed \cite{rodrigo2021double, 10.1145/3387902.3392617}. In \cite{10.1145/3387902.3392617}, the authors propose a method for mapping quantum programs to a modular
quantum architecture based on graph partitioning techniques. However, this approach is only tested on a relatively small and fixed quantum computing multi-core architecture in which the number of cores and qubits per core are both constant (i.e. 10 cores $\times$ 10 qubits per core) irrespective of the width (i.e number of qubits) of the circuit to be executed.  

This paper focuses on the very new field of compilation techniques for scalable multi-core quantum computer architectures with the aim of performing distributed quantum computing. To this purpose, the challenges of mapping quantum algorithms to these modular architectures are discussed in Section II, emphasizing the main differences with single-core mapping methods. In section III, we introduce one of the most recent and advanced works on mapping for modular architectures \cite{10.1145/3387902.3392617}. In Section IV, we further explore the performance of this mapping approach by performing an architectural scalability analysis. Finally, conclusions are presented in Section V.

\begin{figure*}[!h]
    \centering
    \includegraphics[width=\linewidth]{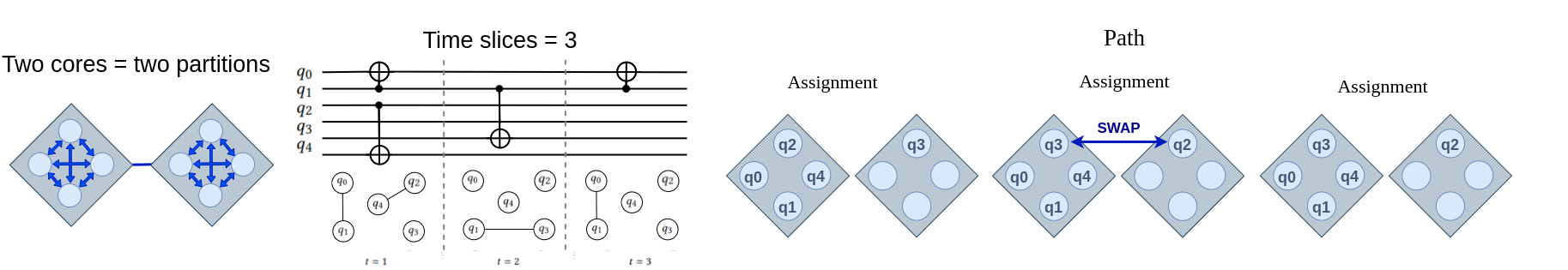}
    \caption{ Example of the quantum circuit mapping technique proposed in~\cite{10.1145/3387902.3392617}. A multi-core quantum architecture composed of 2 cores and 4 qubits per core with all-to-all connectivity is shown on the left. Next, the quantum circuit to be mapped with its respective time slices is presented. Note that each time slice can be represented by a qubit interaction graph in which virtual qubits correspond to the nodes and edges are the interactions between them (i.e. two-qubit gates). For each of the time slices, a valid assignment is returned by using a relaxed version of the so-called Overall Extreme Exchange (rOEE) algorithm. To achieve it, qubits are exchanged between cores by means of SWAP gates, until a valid assignment is found.}

    \label{fig::mapp_exmple}
\end{figure*}

\section{From single-core to multi-core mapping}

Quantum circuit mapping techniques have been developed for single-chip NISQ processors, as part of the compilation process, to deal with their constraints and allow to successfully execute quantum algorithms~\cite{https://doi.org/10.48550/arxiv.2007.01000,mappings, 8382253, 7059001, Venturelli_2018,10.1145/3168822,8980312,10.1145/3297858.3304075}. More precisely, quantum circuit mapping is about transforming hardware-agnostic quantum circuit descriptions into a hardware-compliant version that considers all physical restrictions of a given quantum processor. One of the main constraints in current quantum devices is the reduced connectivity between physical qubits, which usually limits their possible interactions to only nearest-neighbour requiring qubits to be moved to adjacent positions to execute the desired two-qubit operation (e.g. CNOT gate). The circuit mapping procedure consists of different steps: i) \textbf{gate decomposition}, in which gates of the circuit are decomposed into a series of native gates implementable in the quantum processor; ii) \textbf{initial placement} of qubits, where quantum circuit qubits, i.e. virtual qubits, are assigned to the physical qubits of the device. This process helps to minimize the (movement) operations needed in the routing stage; iii) \textbf{routing} of qubits, in which non-neighboring qubits that need to perform a two-quit gate are moved to adjacent physical qubits (which share a connection) by means of, for instance, SWAP gates; and iv) \textbf{scheduling} of operations to leverage parallelism while respecting their dependencies and quantum hardware constraints. An example of the quantum circuit mapping process is shown in Figure~\ref{fig:map_example}.

As mentioned in the previous section, multi-core quantum computing architectures are a promising approach to scale up current single-core quantum computers.
%alleviating the increasing number of errors. 
Existing proposals agree on an architecture based on interconnecting multiple NISQ processors~\cite{QuantumIntranet, https://doi.org/10.48550/arxiv.2201.08861} consisting of tens to hundreds of qubits, increasing the total qubit count without losing that much fidelity and improving isolation. In this architectural design, NISQ processors will be ultimately interconnected through short-range quantum-coherent links and classical links in the form of a so-called `quantum intranet'~\cite{QuantumIntranet}. Quantum coherent links will be responsible for transporting qubits (or quantum states) from core to core, for instance, by means of shuttling o quantum teleportation. Several challenges arise with this new architecture, being the most relevant for this work the need for exchanging quantum information between cores. %Similar to the single-core mapping problem, qubits will need to find a way to interact across cores, but in this case from core to core, whenever they need to interact.
Note that these inter-core communications are more expensive and error prone than those performed in single-core architectures. Therefore, multi-core quantum computing architectures require the development of a new breed of compilation techniques that will have to consider the following fundamental different aspects:

\textbf{Inter-core communication}: Similar to the single-chip case in which qubits need to be adjacent for interacting, qubits placed in different cores cannot directly perform a two-qubit gate. To do so, they have to make use of entanglement-based quantum communication protocols that require the generation of the so-called \textit{Bell pairs} allowing to perform, for instance, remote CNOTs between distant qubits or to teleport quantum states from one core to another~\cite{9334411, rodrigo2022characterizing}. This comes with an overhead of resources needed for creating and distributing entangled pairs. In addition, the entanglement generation is a non-deterministic process making the scheduling task more complex.

\textbf{Not all qubits have the same functionality}: In each of the quantum cores there will be qubits devoted to computation and storage and qubits used for communication. Communication qubits will handle inter-core communications, whereas storage qubits will perform local operations. Mapping techniques will have to include information about qubit `types' and which ones are being used as well as the resources available for communication. Note that the more qubits are dedicated to communication, the higher the number of inter-core communications that can be performed in parallel.

%\newpage
 
\textbf{ A two-step quantum circuit mapping process}: An initial qubit placement and routing should be done at the quantum core level, placing qubits that need to interact on the same core and use efficient routing techniques to reduce inter-communication operations, but also within the quantum processors to reduce the overhead created due to their limited qubit connectivity as in  the single-chip case. 

%by optimally planning and moving qubits among cores. Following it, mapping approaches for single-core processors should be applied separately to each core.

    %\item  The time consumption of qubit movements between cores is not deterministic, requiring a dynamic scheduler.
    
    %\item  The scheduler must have control of the available resources, i.e. which qubits are available and which are occupied performing inter-core movements.
    %\item The initial placement should place the qubits considering the different qubit movement costs and qubits interactions: qubits that will interact must be placed, if possible, on the same core.
    
   % \item The mapping must be performed in two steps: first, optimally planning and moving qubits among cores; second, mapping approaches for single-core processors should be applied separately to each core. 
%\end{itemize}

%There are several mapping solutions for single-core NISQ devices~\cite{https://doi.org/10.48550/arxiv.2007.01000,mappings, 8382253, 7059001, Wille2016UsingD, Venturelli_2018}. However, only a mapping approach for mapping quantum algorithms in multi-core quantum architectures~\cite{10.1145/3387902.3392617}, has been proposed so far, which will be discussed in the next section. 

Multi-core or modular architectures for scaling up quantum devices share a lot of similarities with the quantum networks that are being deployed for a future quantum internet~\cite{van2016path,https://doi.org/10.48550/arxiv.2201.08825,8910635}. The main difference resides in the fact that communications links, instead of being short-range, are long-range~\cite{Kozlowski_2019}, resulting in the need for a more complicated infrastructure to move qubits between quantum devices, i.e. quantum repeaters. Due to this quantum network infrastructure, moving qubits among devices would be more complex, needing to perform entanglement swaps~\cite{9334411}, which increases latency considerably as its duration grows exponentially with the distance between devices. One possible application of quantum networks is to perform distributed quantum computing, for which compilation techniques have been already proposed~\cite{9334411,https://doi.org/10.48550/arxiv.2112.14139,Cicconetti_2022}. However, not so much attention has been paid so far to the development of compilers for multi-core quantum computing architectures \cite{rodrigo2021double, 10.1145/3387902.3392617}. In the next sections, we  will focus on the mapping technique proposed in~\cite{10.1145/3387902.3392617}. % and further analyse its performance by performing an architectural scalability analysis. 

\section{Quantum circuit partitioning for distributed quantum computing}

%           ---> !!! FIGURES IN THE MAIN FILE FOR EASY ORDERING !!! <----
%Multi-core or modular quantum architectures %are a new technology~\cite{Monroe_2014, https://doi.org/10.48550/arxiv.2201.08825, https://doi.org/10.48550/arxiv.2201.08861,https://doi.org/10.48550/arxiv.2210.10921} proposed to overcome the scalability problem encountered in current single-core or monolithic processors. This new technology
%introduce new challenges, such as the movement of qubits between cores. As stated in the last section, these movements are costly and should be avoided to the extent possible. Optimal compilation or mapping is required to reduce these inter-core movements. So far, only one mapping technique has been proposed for multi-core or modular quantum computing architectures, presented 

In~\cite{10.1145/3387902.3392617} a technique for mapping quantum algorithms on multi-core architectures based on graph partitioning has been proposed. The goal is to place qubits in the different quantum processors such that inter-core movements are minimized. An illustrative example of this mapping technique is shown in Figure~\ref{fig::mapp_exmple}. %For a more detailed explanation we refer the reader to ~\cite{10.1145/3387902.3392617}.  
Note that in the proposal presented in~\cite{10.1145/3387902.3392617} the following assumptions that simplify the quantum circuit mapping problem are made: i) all-to-all connectivity between cores and among physical qubits within the cores. This means that there is no need for qubit routing inside the core, nor for optimal initial placement. Regarding inter-core routing and qubit placement at the core level, all qubits are at a one-hop distance, and therefore when two qubits have to interact and cannot be placed from the beginning on the same quantum core it is enough to place one of them on any other core; ii) SWAP operations (i.e. exchange of quantum states) are used for inter-core communication that makes simpler the management of resources as it not required to check if there is space (i.e. qubits that do not have any information) for exchanging qubits between cores; iii) only a fixed modular architecture is considered consisting of 10 cores with 10 qubits per core, which is not enough for analyzing the performance of the quantum circuit mapper. In the following section, different architectures will be used to further analyze this proposed mapping procedure.

\section{Results}

%\begin{figure*}[!h] 
%     \centering
%    \subfloat[]{
 %           \includegraphics[width=0.3\textwidth]{fig/results/nCom_Strong_Comp_cucc.png}\label{fig:sub5}
 %   }
  %  \subfloat[]{
  %      \includegraphics[width=0.3\textwidth]{fig/results/time_fix_var_Strong_Comp_cucc.png}\label{fig:sub3}
   % }
    %\subfloat[]{
    %    \includegraphics[width=0.3\textwidth]{fig/results/weak_scaling.png}\label{fig:sub4}
    %}
    %\caption{Experiments results for different architectures. The comparison between a fixed number of qubits per core and a variable number of qubits per core is shown in the first and second rows, focusing on the non-local communications (SWAPs) and execution time, respectively. The weak and strong scaling results are shown in the last row.}
    %\label{fig:bench_comp}
%\end{figure*}

\begin{figure}[!h] 
     \centering
    \subfloat[]{
        \includegraphics[width=0.49\columnwidth]{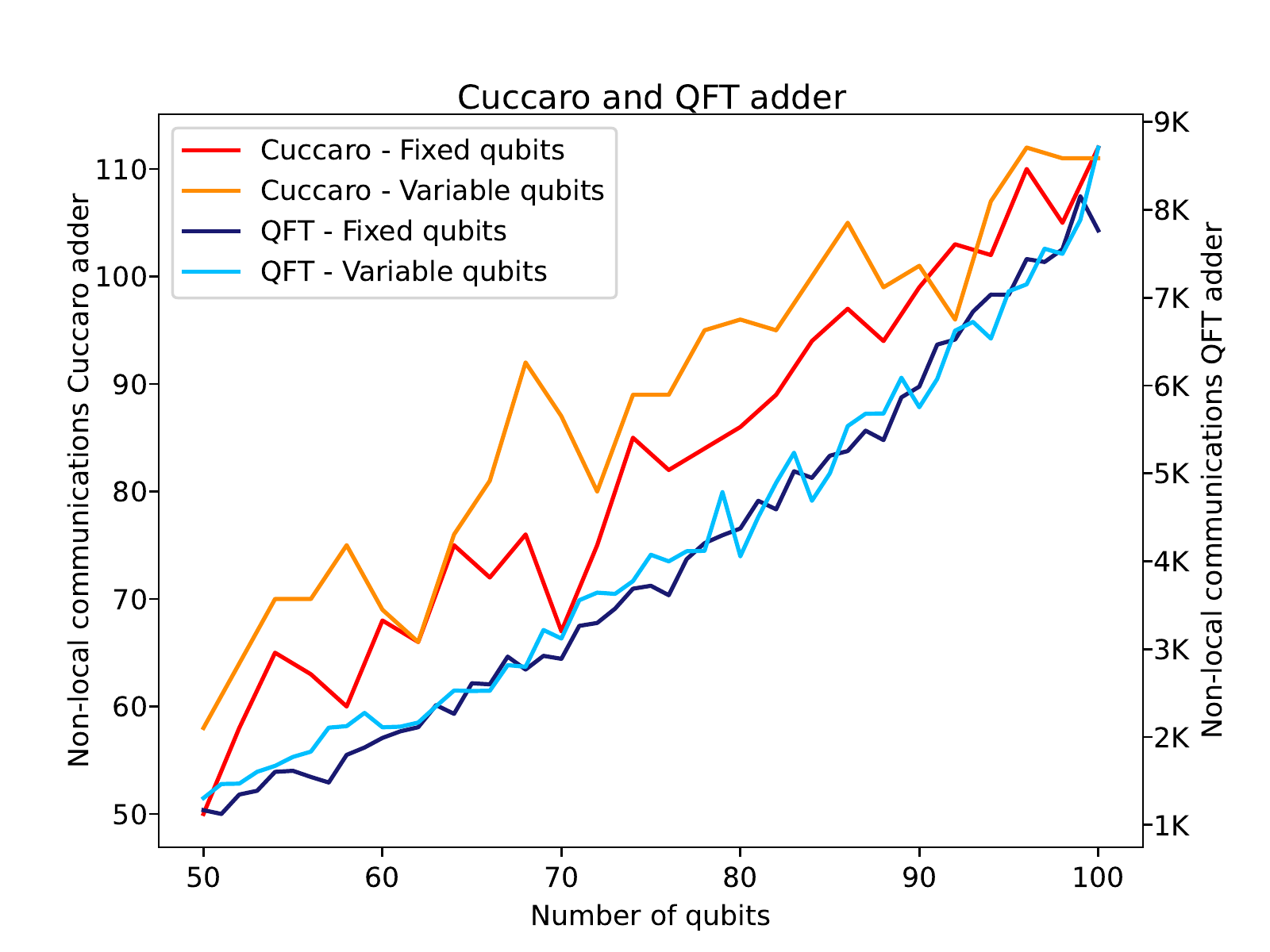}\label{fig:sub1}
        }
    \subfloat[]{
        \includegraphics[width=0.49\columnwidth]{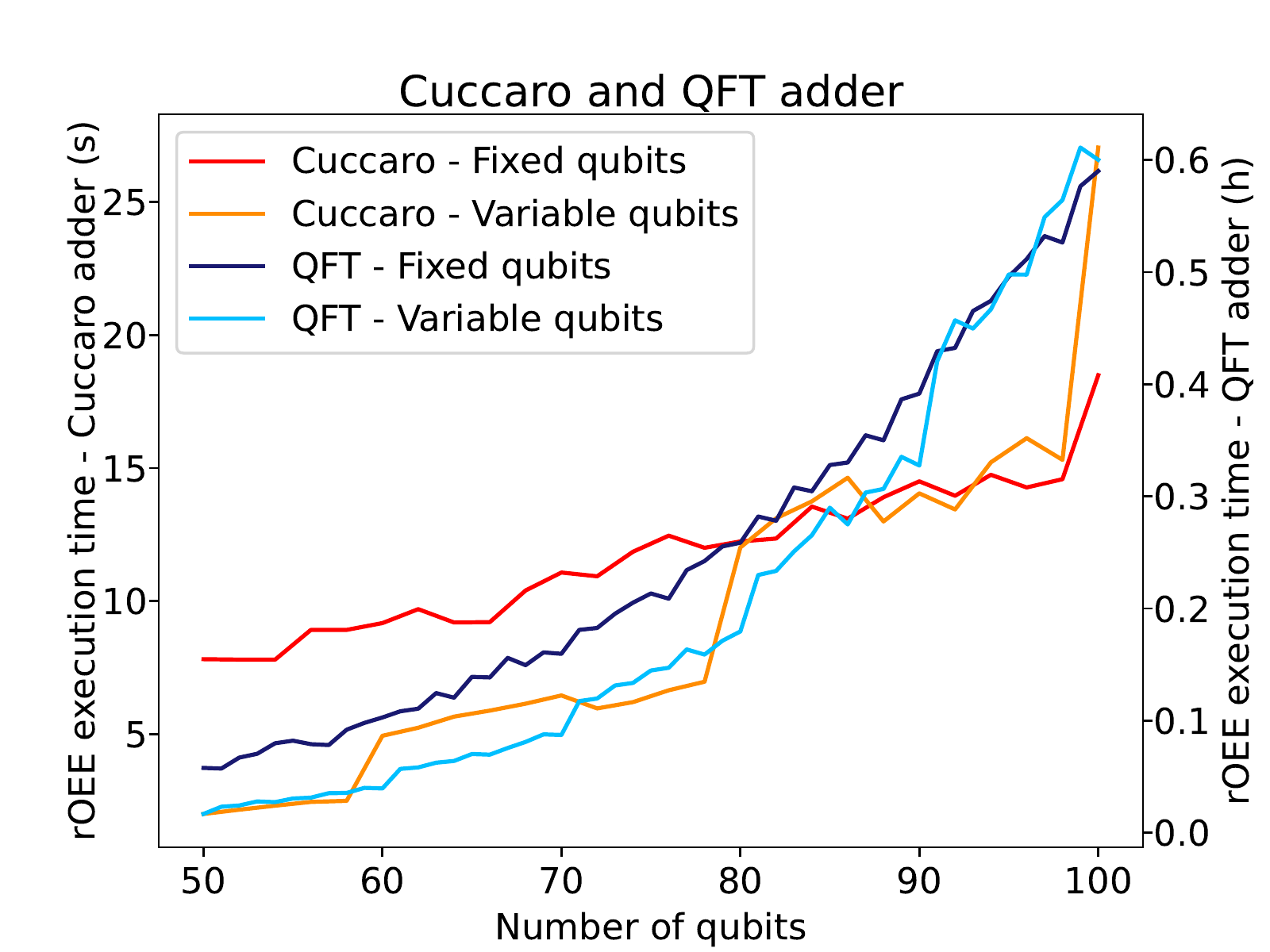}\label{fig:sub2}
    }\\
    \vspace{-9pt}
    \subfloat[]{
           \includegraphics[width=0.49\columnwidth]{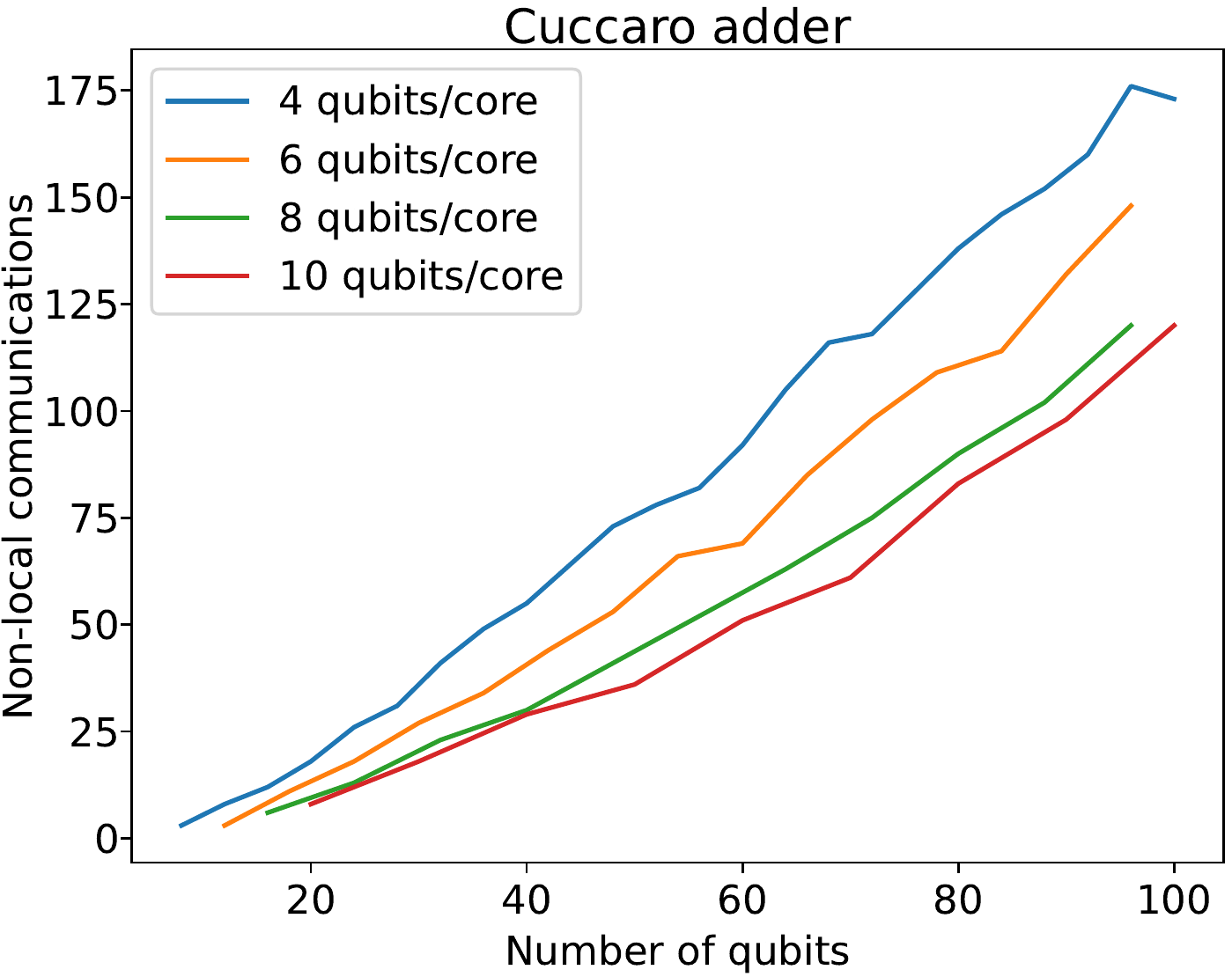}\label{fig:sub5}
    }
    \subfloat[]{
      \includegraphics[width=0.49\columnwidth]{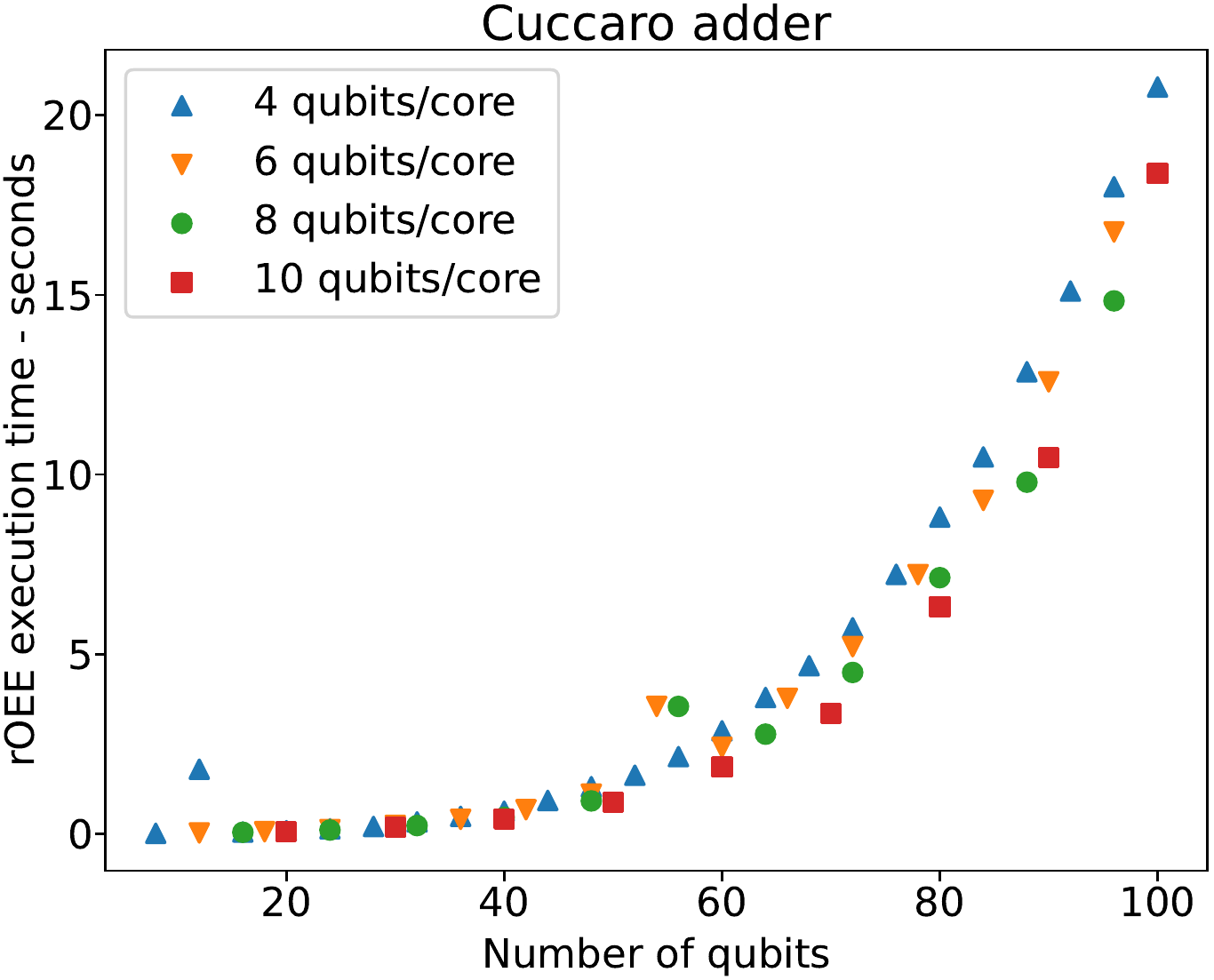}\label{fig:sub3}
    }\\
    \vspace{-9pt}
    \subfloat[]{
       \includegraphics[width=0.49\columnwidth]{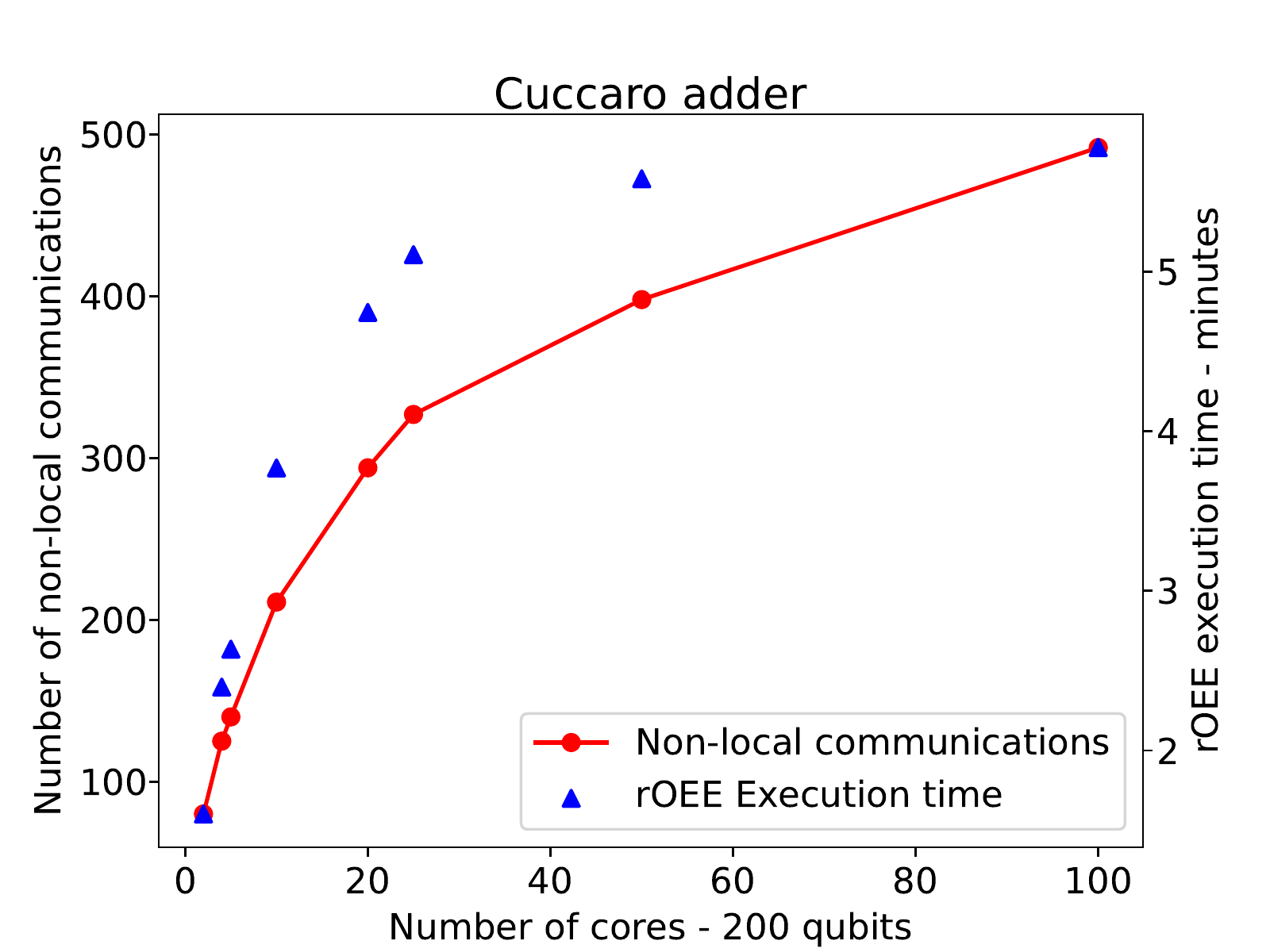}\label{fig:sub4}
    }
    \caption{Non-local communications (SWAPs) and execution time for different multi-core architectures. (a) and (b) for a fixed and a variable number of qubits per core. (c) and (d) when a strong scaling of the architecture is performed. (e) Weak scaling of the multi-core architecture.}
    \label{fig:bench_comp}
\end{figure}

%           ---> !!! FIGURES IN THE MAIN FILE FOR EASY ORDERING !!! <----

%The work presented in~\cite{10.1145/3387902.3392617} is, so far, the only mapping technique for multi-core or modular quantum architectures. Their results are based on an architecture of 10 cores and 10 qubits per core with all-to-all connectivity.

As previously mentioned, in this work we further analyze the performance of the quantum circuit mapping approach in ~\cite{10.1145/3387902.3392617} by considering different multi-core architecture designs with  still all-to-all qubit and cores connectivity. For this purpose, several quantum benchmarks have been used. %~\cite{thesis}. 
In this paper, results for the Cuccaro and the QFT adder are presented as they have very dissimilar circuit characteristics. The Cuccaro adder is a well parallelizable and easy to scale algorithm with a low number of two-qubit gates and circuit depth. In contrast, the QFT adder is a more sequential algorithm with a huge number of two-qubit gates and large depth. In addition, two performance metrics are used: the number of non-local communications (i.e. inter-core movements) and the execution time (i.e. time it takes to calculate all valid assignments).

%\textcolor{red}{Proposal: 1)combine figures (a) and (b) on the same graph. 2) combine figures (d) and (e) on the same graph. 3) remove (c) and (f). 4) Leave strong and week scaling but not to use quantum volume unless it is necessary.}

 Figure~\ref{fig:sub1} compares how the mapping algorithm behaves when a fixed and a variable number of qubits per core is assumed, both architectures counting with ten cores. In other words, in the first case cores always consist of 10 qubits per core independently of the circuit width, whereas in the second case, the minimum (even) number of qubits per core is used based on the algorithm requirements.    %It should be noted that when a variable number of qubits per core is considered, the number of qubits should always be even since an odd number of qubits could cause a failure in the mapping procedure. 
Similar behavior with respect to non-local communications can be observed for a fixed and variable number of qubits for both quantum circuits. However, note that the difference between both cases is much more pronounced for the Cuccaro adder due to its circuit characteristics. This means that the relevance of the number of physical qubits in the architecture regarding non-local communications depends on the algorithm to be executed. In contrast, the total number of physical qubits in the architecture is crucial for the rOEE runtime. As shown in Figure~\ref{fig:sub2}, a lower execution time is required for the variable case. The reason is that the rOEE algorithm computes over physical qubits to find a valid assignment, and therefore the more physical qubits, the more iterations are needed, increasing the execution time. Furthermore, note the large difference in non-local communications as well as in execution time between the Cuccaro and the QFT adder.

%When the number of physical qubits is fixed the execution time scales linearly with the logical qubits. However, when the number of physical qubits increases with the logical qubits, the execution time drastically increases with the number of physical qubits. The rOEE computes over physical qubits to find a valid assignment, the more physical qubits, the more operations it performs, increasing the execution time,  leading to the conclusion that the number of physical qubits that an architecture has may have a limit due to the execution time. This experiment has been repeated for a larger number of physical and logical qubits, resulting in the same behavior.

In addition, two more architectural scalability experiments have been performed, named \textit{weak and strong} scaling. In weak scaling, the total number of physical qubits is fixed, whereas the number of cores and qubits per core varies, increasing the number of cores while decreasing the number of qubits per core. In Figure~\ref{fig:sub4} the weak scaling results are shown; both, the rOEE runtime as well as the non-local communications increase when more cores are added to the architecture. The more cores and fewer qubits per core, the more computations will be performed until the rOEE algorithm finds a valid assignment and the higher the inter-core movements are. %Note that the runtime rises more when the number of cores increases than the non-local communications

In strong scaling, the number of qubits per core is fixed but we increase the number of cores and therefore the total number of qubits in the device. Four different architectures have been used with 4, 6, 8, and 10 qubits per core, starting with 2 cores and increasing them until a total number of 100 qubits is reached. As shown in Figures~\ref{fig:sub5} and~\ref{fig:sub3},  non-local communications increase as more cores are added. Moreover, on architectures with fewer qubits per core, a higher number of non-local communications is observed due to higher constraints to find a valid assignment. The execution time in relation to the total number of qubits is similar for  all four cases since, as mentioned before, the most crucial parameter concerning execution time is the total number of physical qubits and not how they are distributed. 

\section{Conclusions}
Multi-core or modular quantum computing architectures are a promising approach to overcome the scaling difficulties encountered in monolithic or single-core quantum processors. However, this new architectural design comes with a set of challenges such as qubit interactions across cores. Inter-core communications can be minimized through the process of mapping, as proposed  in~\cite{10.1145/3387902.3392617}. In this paper, we have further analyzed the performance of this quantum circuit mapping technique by performing several experiments in which different architectures with all-to-all connectivity are considered. The most important findings can be summarized as follows: i) Non-local communications and execution time increase with the circuit width. ii) The number of physical qubits is the most important factor regarding execution time and therefore, using a variable number of qubits per core is more efficient. iii) The higher the number of cores, the longer the execution time and the higher the non-local communications. Note also, that the performance highly depends on the quantum algorithm to be executed. 

%Although the proposed technique operates successfully, it has an important constraint: the number of physical qubits is limited due to the long execution time. An approach to overcome it would be adapting the number of physical qubits used to the logical qubits the quantum circuit has. However, further research will be necessary to overcome all challenges multi-core architectures present.

\section*{Acknowledgments}

We acknowledge support from EU, grant HORIZON-ERC-101042080 (S.A.) and grant HORIZON-EIC-2022-PATHFINDEROPEN-01-101099697 (S.A., E.A and C.G.A.), from project PRG 946 funded by the Estonian Research Council (A.O.), from MICIIN and European ERDF under grant PID2021-123627OB-C51 (C.G.A.) and from MICIIN with funding from European Union NextGenerationEU(PRTR-C17.I1) and by Generalitat de Catalunya (E.A.)

\bibliographystyle{ieeetr}
\bibliography{bib/main}
%\printbibliography

\end{document}